\documentclass[9.5pt]{article}
\usepackage[margin=0.7in]{geometry}

\usepackage{verbatim}
\usepackage{amsmath}
\usepackage{amssymb}
\usepackage{multirow}
\usepackage{apacite}

\usepackage{booktabs}

\usepackage{xcolor}
\usepackage{tabu}
\usepackage{float}
\usepackage{graphicx}

\usepackage[utf8]{inputenc}
\usepackage{gensymb}
\usepackage[section]{placeins}
\usepackage[sort&compress,sectionbib]{natbib}

\usepackage{lineno}
\usepackage{authblk}

\usepackage{titlesec}
\titleformat*{\section}{\large\bfseries}
\titleformat*{\subsection}{\normalsize\bfseries}

\usepackage{setspace}
\usepackage{etoolbox}
\preto\longtable{\par\singlespacing}

\usepackage{hyperref}
\hypersetup{
	colorlinks = true,
	linkcolor = {red},
	citecolor = {blue},
	linkbordercolor = {white},
	citebordercolor = {purple}
}

\title{Lessons learned from NASA's DART impact about disrupting rubble-pile asteroids}

\author[1]{S. D. Raducan}
\author[1]{M. Jutzi}
\author[2]{C. C. Merrill} 
\author[3,4]{P. Michel} 
\author[5]{Y. Zhang}
\author[6]{M. Hirabayashi} 
\author[7]{A. Mainzer}

\affil[1]{Space Research and Planetary Sciences, Physikalisches Institut, University of Bern, Bern, Switzerland}
\affil[2]{Sibley School of Mechanical and Aerospace Engineering, Cornell University, Ithaca, NY, USA}
\affil[3]{Universit\'e C\^ote d'Azur, Observatoire de la C\^ote d'Azur, CNRS, Laboratoire Lagrange, Nice, France}
\affil[4]{The University of Tokyo, Department of Systems Innovation, School of Engineering, Tokyo, Japan}
\affil[5]{Department of Climate and Space Sciences and Engineering, University of Michigan, Ann Arbor, MI 48109, USA}
\affil[6]{Georgia Institute of Technology, Atlanta, GA, USA}
\affil[7]{University of Arizona, Lunar and Planetary Laboratory, Tucson, AZ, USA}
\date{}
\begin{document}


\maketitle

\maketitle

\begin{abstract}
We present a series of numerical simulations using a shock physics smoothed particle hydrodynamics (SPH) code, investigating energetic impacts on small celestial bodies characterised by diverse internal structures, ranging from weak and homogeneous compositions to rubble-pile structures with varying boulder volume packing. Our findings reveal that the internal structure of these rubble-pile bodies significantly influences the impact outcomes. Specifically, we observe that the same impact energy can either catastrophically disrupt a target with a low boulder packing  ($\lesssim$\,30 vol\%), or result in the ejection of only a small fraction of material from a target with the same mass but high boulder packing ($\gtrsim$\,40 vol\%). This finding highlights the pivotal role played by the rubble-pile structure, effectively acting as a bulk shear strength, which governs the size and behaviour of the resulting impact. Consequently, understanding and characterising the internal structure of asteroids will be of paramount importance for any future efforts to deflect or disrupt an asteroid on a collision course with Earth.
\end{abstract}

\section{Introduction}


In a significant milestone for human history, NASA's DART (Double Asteroid Redirection Test) spacecraft impacted the asteroid Dimorphos, the secondary of the Didymos asteroid binary system, on September 26th, 2022 (UTC) \citep{Rivkin2021,Daly2023,Chabot2023}. The impact resulted in an orbital change of approximately 33 minutes of Dimorphos around its primary, Didymos \citep{Thomas2023}. This groundbreaking achievement demonstrated the capability to redirect the trajectory of a potentially hazardous asteroid, paving the way for future efforts to protect our planet from potential threats. The DART spacecraft with a mass of 580 kg collided with Dimorphos at 6.15 km/s, hitting the target within 25 m of the centre of figure of the asteroid and at an incidence angle of only 16.7 $\pm$ 7.4$^\circ$ from the average surface normal. This angle was calculated based on a 1.5-meter radius around the impact point, using data from the global digital terrain model (DTM) of the impact site \citep{Daly2023}.

Dimorphos is an oblate ellipsoid, with $\approx$\,87.90$\times$86.96$\times$57.16 m \citep{Daly2023}, having an equivalent volume of a $\approx$\,150 m sphere. ESA's Hera mission will visit the Didymos binary system in late 2026 for rendezvous investigations.

To date, we have discovered more than 32,000 objects in the Near-Earth Object (NEO) population\footnote{NEA discovery statistics as of September 2023 (cneos.jpl.nasa.gov).}. Only approximately one-third of these objects exceed a size of 140 meters. However, it is noteworthy that less than half of the asteroids falling within the size range of 20 to 140 meters have been located to date \citep{Harris2021}, many of which are not well catalogued. The prediction is that the impact frequency of objects in this size range is higher due to the increasing number of smaller objects. Ongoing initiatives are underway to identify and monitor asteroids within this critical size range. This casts the significance of better characterising how such small asteroids contribute to the hazards on the Earth.


To put it into perspective, consider the Chelyabinsk event in February 2013 \citep{Popova2013}. This incident involved a relatively small asteroid, approximately 20 meters in diameter \citep{Artemieva2016}, which entered Earth's atmosphere and exploded at an altitude of 27 km near Chelyabinsk, Russia. Despite its modest size, the explosion caused extensive damage, primarily to buildings in the area due to shock wave propagation. The only other large meteoroid airbust recorded in the last century occurred on June 30th, 1908 over the Tunguska forest in Siberia (Russia). The Tunguska event produced even more extensive damage, flattening approximately 2000 square km of forest \citep{Artemieva2016}. The event is believed to have been caused by a 30 to 50 m diameter asteroid \citep{Artemieva2016} that exploded 6-10 km in the atmosphere. 

Upon detecting a potentially hazardous asteroid heading toward Earth, the immediate priority lies in determining our course of action. This pivotal decision relies on a thorough analysis of the asteroid's size, composition, and trajectory \citep[e.g.,][]{Rumpf2020}. Different asteroids may require different techniques and energies to deflect them, and a deflection strategy is usually favoured over disruption, where at least half of the target's original mass is ejected and leads to the creation of a cloud of potentially hazardous fragments with uncertain size.

With DART's success, we now know that a kinetic impactor is a viable option to deflect a potentially hazardous asteroid, as long as we do it long enough in advance. But the DART mission was just the first kinetic impactor test, and missions to impact bodies smaller than Dimorphos are being designed \citep[e.g.,][]{Merrill2024} and planned with launch dates even as early as 2025 \citep[e.g.,][]{Wang2023}. However, deflection missions require detailed planning, and we need to be able to answer the question ``What is the smallest asteroid size that is feasible to be deflected by a kinetic impactor, without disrupting it?". Here, we use numerical models of asteroid impacts to address this question. We leverage the invaluable data from the DART impact and extrapolate it to rubble-pile asteroids of different sizes. 

\section{Insights gained from the DART impact about the mechanical properties of Dimorphos}

Prior to the impact, there was a large uncertainty regarding the mechanical properties of both Didymos and Dimorphos, particularly concerning their structure and surface cohesion \citep{Raducan2019, Raducan2020, Stickle2022}. Geological assessment of the two asteroids, based on images captured by the onboard DRACO camera during the DART approach, suggests that both asteroids are rubble-pile objects \citep{Barnouin2023}, i.e., accumulations of debris held together primarily by self-gravity and/or small cohesive forces \citep{Richardson2022, Scheeres2010, Walsh2008}. \cite{Barnouin2023} concluded that the surface of Dimorphos has a cohesion of at most 0.03 Pa and estimated a friction angle of 35 degrees, however, a stiffer interior is not excluded.

These results are also supported by numerical simulations of the DART impact outcome. \cite{Raducan2023} compared simulation results from an extensive set of runs, systematically varying the mechanical properties of the target (e.g., interparticle cohesion, coefficient of internal friction, bulk porosity) and the boulder size-frequency distribution (SFD), with the observed deflection efficiency \citep{Cheng2023}, the shape and morphology of the ejecta cone \citep{Li2023, Dotto2023}, and the estimated amount of ejecta mass \citep{Graykowski2023}. They found that in order to reproduce all the observables, Dimorphos must be a rubble pile, with little or no surface cohesion ($Y<$ a few Pa) and a low packing of boulders on the surface. A low boulder packing is defined as less than $\approx$ 30\% of the volume occupied by boulders larger than 2.5 m. Based on boulder SFD and boulder shape \citep{robin2023}, they also estimate that the surface macroporosity is about 35\%. 

Numerical predictions of the DART impact outcome \citep{Raducan2023} together with observations of the secondary lightcurve deduced from high-quality photometric observations [Pravec et al., in prep.] strongly suggest that the DART impact occurred in the subcatastrophic impact regime (i.e., a regime between cratering and catastrophic disruption) and that the impact caused significant reshaping of Dimorphos, as opposed to an impact crater \citep{Raducan2022, Raducan2022d}. The models suggest a change in the $a_s/b_s$ ratio of the ellipsoidal axes from 1.06 \citep{Daly2023} to up to 1.2 \citep{Raducan2023}, while observations suggest $a_s/b_s \approx 1.3\pm0.1$, if the body is currently in or close to a tidally locked state with minimal libration amplitude and an approximately zero obliquity [Pravec et al., in prep.]. 

The ESA's Hera mission will investigate the Didymos system in early 2027 \citep{Michel2022} and will allow us to measure in detail the DART impact outcome, i.e., the crater's size or global reshaping of Dimorphos, as well as Dimorphos' internal properties, offering complete documentation of the DART impact experiment for impact code validation.

\section{Previous studies}

The catastrophic disruption threshold, $Q^*_D$, is the specific impact energy required to disperse half of the initial target mass. Historically, this threshold has been estimated through laboratory and numerical hydrocode experiments. While these methods have been valuable, they face limitations when applied to small bodies in the 10--100 m range, where the intricate interplay of material strength, friction, porosity, and self-gravitation governs outcomes. Laboratory experiments become impractical at this scale, leading to the increased use of numerical approaches. For example, \citet{Benz1999} employed a smoothed particle hydrodynamics (SPH) code to model the fracturing of basalt and icy bodies. They integrated a tensile fragmentation model but did not account for target porosity or heterogeneities. \citet{Leinhardt2009} used the CTH code \citep{McGlaun1990} to study the dependence of $Q^*_D$ on the limiting yield strength of the target, focusing on homogeneous solid objects. In \citet{Jutzi2010}, the effects of target microporosity on $Q^*_D$ was explored using a sub-resolution porosity model ($P-\alpha$ model) \citep{Jutzi2008}. Subsequent studies by \citet{Jutzi2015} and \citet{Arakawa2022} also explored the effects of target friction and cohesion, employing a combination of laboratory experiments and numerical simulations.

Most of the previous studies used internally homogeneous objects. \citet{Benavidez2012,Benavidez2018} investigated collisions among ``rubble-pile'' bodies using the SPH code by \cite{Benz1994,Benz1995}. However, these simulations did not account for friction between the boulders/grains, leading to significantly low $Q^*_D$, as discussed in \citet{Jutzi2015}. The collisional strength of small (10 -- 100 m) rubble-pile asteroids has yet to be systematically studied using realistic material properties, and recent work showed that the presence of macroscopic boulders has a large influence on the impact response of these objects \citep{Raducan2022d}. 
The novelty of the work presented in this study lies in its use of material models and parameters calibrated to an actual asteroid impact, marking the first instance of such an approach. We have used the DART impact experiment on asteroid Dimorphos to determine the optimal material models and simulation parameters for the Bern SPH code, including material strength, porosity, and boulder configuration.

\section{Numerical models}

We use the Bern smoothed particle hydrodynamics (SPH) impact code \citep{Benz1995, Jutzi2008, Jutzi2015}, to numerically model impacts of varying specific energies, over a range of assumed sets of material properties and interior structures for the target. From our simulations, we compute the size of the largest remnant and compute the catastrophic disruption threshold, $Q^*_D$. 

Bern SPH is a shock physics code originally developed by \cite{Benz1994, Benz1995} to model the collisional fragmentation of rocky bodies and was later parallelised \citep{Nyffeler2004} and further extended by \citep{Jutzi2008} and \citep{Jutzi2015} to model porous and granular materials. The most recent version of the code includes a tensile fracture model \citep{Benz1995}, a porosity model based on the $P-\alpha$ model \citep{Jutzi2008}, pressure-dependent strength models \citep{Jutzi2015}, and self gravity. The Bern SPH code has been validated in a number of studies \citep[e.g.,][]{Jutzi2009, Jutzi2015, Ormo2022} and benchmarks against other codes \citep[e.g.,][]{Luther2022}. 

\subsection{Rubble-pile models}
\begin{figure}[ht]
\centering
\includegraphics[width=\linewidth]{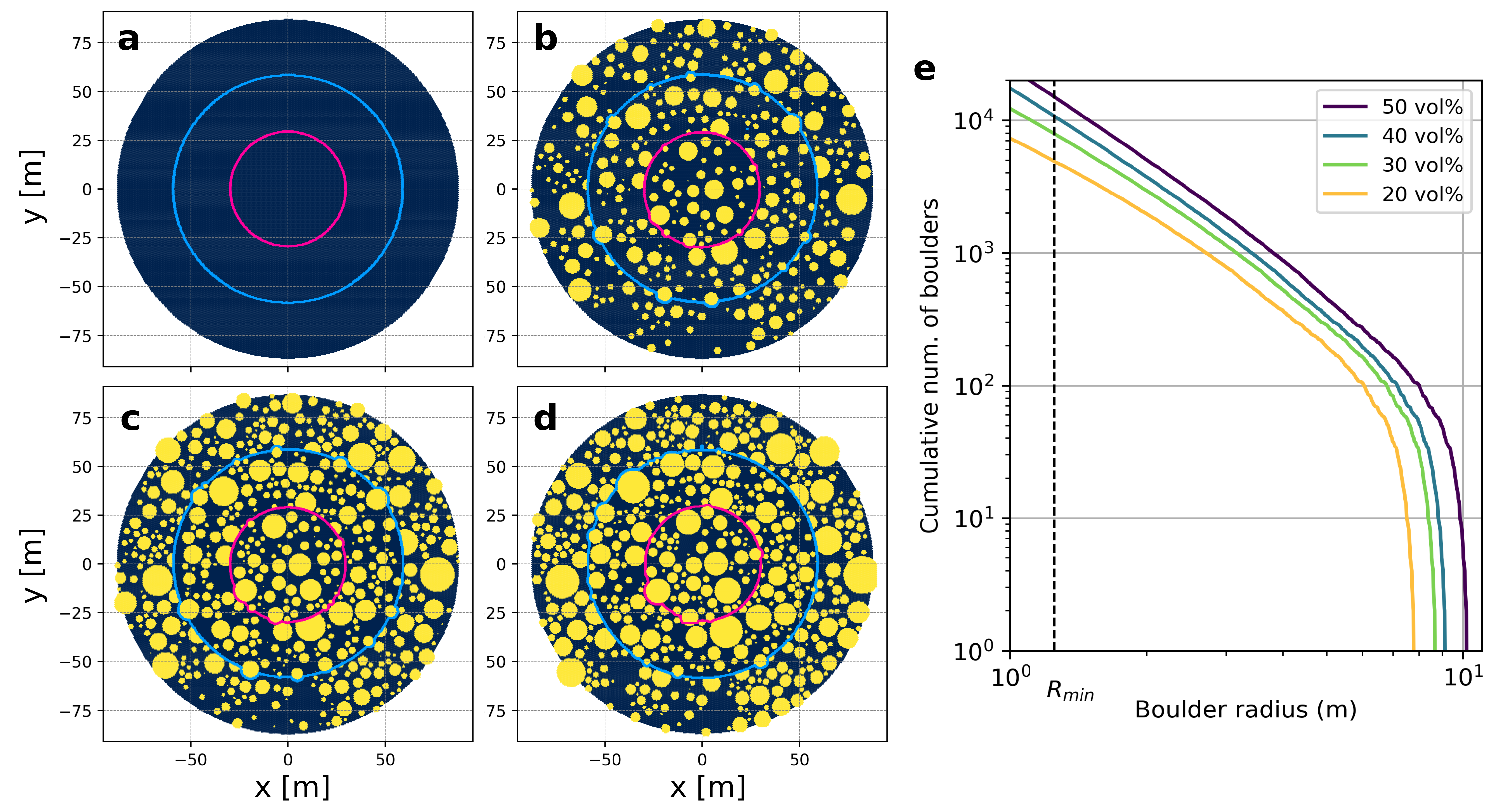}
\caption{Cross-sections through the targets used in the SPH simulations, with different boulder packings: a) 0 vol\%, b) 30 vol\%, c) 40 vol\% and d) 50 vol \%. In each panel, the outlines show the three target sizes simulated: 87.90$\times$86.96$\times$57.16 m;  58.60$\times$57.97$\times$38.11 m; and 29.30$\times$28.99$\times$19.05 m. In all simulations, the impact is vertical, along the $y$ axis. e) Boulder size-frequency distributions (SFD) for each boulder packing studied here. The 20 vol\% packing SFD is plotted for comparison. Due to resolution constraints, boulders smaller than $R_{\text{min}}$ = 2.5 m are not explicitly modelled; instead, they are included in the matrix material and treated as a continuum.}
\label{fig:boulders}
\end{figure}

We model kinetic impacts into rubble-pile ellipsoidal targets composed of spherical boulders with different distributions embedded into a matrix material. We used the N-body code pkdgrav \citep{Richardson2000} to generate rubble-pile targets, and we used Dimorphos' surface boulder size-frequency distribution (SFD) \citep{Pajola2023} to generate boulder dimensions. To explore a large possible range of boulder mass fractions, we removed some of these boulders from pkdgrav output when we built our SPH models. In addition to a homogeneous target scenario (0 vol\%; Fig.~\ref{fig:boulders}a), we define four different boulder distributions, with $\approx$ 20, 30, 40 and 50\% of the target volume occupied by boulders (Fig.~\ref{fig:boulders}). Boulders smaller than $R_{min}$ = 1.25 m (2.5 m in diameter) are removed from the SFD due to their size being too small to be resolved individually (Fig.~\ref{fig:boulders}c). We assume they are part of the matrix material used to fill the voids between the larger boulders. From the initial pkdgrav boulder re-accumulation, we cut out oblate ellipsoidal targets with three different sizes: a) 87.90$\times$86.96$\times$57.16 m, corresponding to Dimorphos's shape from \cite{Daly2023}; b) 58.60$\times$57.97$\times$38.11 m ($\approx$ 35\% smaller than Dimorphos); and c) 29.30$\times$28.99$\times$19.05 m ($\approx$ 65\% smaller than Dimorphos). Our targets have volume-equivalent diameters of $\approx$ 150 m ($R$ $\approx$ 75 m), $\approx$ 100 m ($R$ $\approx$ 50 m), and $\approx$ 50 m ($R$ $\approx$ 25 m). When we constructed our targets, we aimed to keep a similar target mass for a specific target size, irrespective of boulder packing. Consequently, the target mass for each size varies by only approximately 10\% with different boulder packing densities.

\subsection{Material model}
To model the rubble piles, we assigned material properties that \cite{Raducan2023} and \cite{Barnouin2023} found to be the best fit for the surface mechanical properties of Dimorphos. We modelled both the boulders and the matrix material using the Tillotson equation of state (EoS) for basalt, with modified initial grain densities of $\rho_g$ = 3200 kg$/$m$^3$. The bulk porosity of Dimorphos results from a combination of macroporosity found between individual boulders, as well as microporosity within the boulders themselves. Based on analysis of the reflectance spectra of Didymos, the best matching meteorite analogues are the L/LL ordinary chondrites \citep{deLeon2006, Dunn2013, Ieva2022}. These meteorites have grain densities of $\approx$\,3200–3600 kg$/$m$^3$, and low microporosities of $\approx$\,8–-10\% \citep{Flynn2018}. Therefore, in our simulations, the initial microporosity within boulders was fixed at 10\%. The initial porosity of the matrix (macroporosity + microporosity) was fixed at 45\%, as calculated by \citep{Raducan2023}. The porosity in both the boulders and the matrix was modelled using the $P-\alpha$ porosity compaction model \citep{Jutzi2008}, with a single power-law slope, defined by the solid pressure, $P_s$, elastic pressure, $P_e$, exponent, $n$, and initial distension, $\alpha_0$:
\begin{equation}
  \alpha(P)=\begin{cases}
    1, & \text{if $P_s<P$}\\
    (\alpha_0-1)\left(\frac{P_s-P}{P_s-P_e}\right)^{n}+1, & \text{otherwise.}
  \end{cases}
\end{equation}

The input parameters for the matrix and boulder materials are summarised in Table~\ref{table:model_parameters}. 

In all of our simulations, the boulders, which are represented explicitly, are modelled using a tensile strength and fracture model as described in \citep{Jutzi2008, Jutzi2015}, with parameters corresponding to a tensile strength of $Y_T$ = 10 MPa. While the individual boulders possess significant strength, the inter-particle cohesion is considered to be negligible ($Y_0$ $\approx$ 0 Pa) \citep{Barnouin2023}. We model this matrix material using a simple pressure-dependent strength model \citep{Collins2004, Lundborg1967}, in which the strength asymptotes to a certain shear strength at high pressures. For $Y_0$ = 0 Pa, the Lundborg (LUND) strength model describes the yield strength as:
\begin{equation}
    Y = \frac{fP}{1 + f P/(Y_{dm})}
\end{equation}
where $P$ is pressure, $f$ is the coefficient of internal friction, and $Y_{dm}$ is the limiting strength at high pressure.

The projectile was modelled as an underdense (1000 kg/m$^3$) aluminium sphere and the impact velocity was kept constant at 6 km/s. The projectile hit the target at $x$ = 0, along the y-direction (Fig.~\ref{fig:boulders}). We investigate two impact angles: vertical (0$^\circ$ from the surface normal) and oblique, 45$^\circ$ impact angle. To vary the specific impact energy, we vary the mass of the projectile between 200 kg and 5 $\times$ 10$^5$ kg.

\begin{table}[ht]
 \centering 
	\caption{Material model parameters for impact simulations into Dimorphos analogues.}
    \vspace{0.3cm}
	\begin{tabular}{l@{\hskip 0.1in}l@{\hskip 0.1in}l@{\hskip 0.1in}l}
    Description        & Impactor & Matrix material & Boulders  \\
    \hline
    Material             & Aluminium & Basalt & Basalt \\
	\hline
	Equation of state        & Tillotson$^{a}$ & Tillotson$^b$ & Tillotson$^b$ \\
        Initial bulk modulus, $A$ (GPa) & 7.5 & 26.7 & 26.7 \\
        Fast-integration bulk modulus (MPa) & 1.0 & 1.0 & 1.0 \\
        Grain density, $\rho_g$ (kg/m$^3$) & 1000  & 3200 & 3200 \\
	Strength model           & von Mises & LUND$^c$ & LUND$^c$ \\
	\hline
	LUND strength parameters$^c$ \\
	Damage strength at zero pressure, Y$_0$ (Pa)   & --   & 0  & $1\times10^8$ \\
	Strength at infinite pressure, Y$_{dm}$ (GPa) & 0.34  & 3.5  & 3.5    \\
	Internal friction coefficient (damaged), $f$    & --  & 0.55 & 0.8\\
	\hline
	Porosity model parameters ($P-\alpha$)$^d$           \\ 
	Initial porosity, $\phi_0$             & -- & 45\% & 10\% \\
	Initial distension, $\alpha_0$         & -- & 1.80 & 1.15 \\
	$P_s$ (GPa)                            & -- & 1.0 & 2.0 \\
	$P_e$ (MPa)                            & -- & 1.0 & 1.0  \\
	$n$                                    & -- & 2 & 2  \\
	\hline
    \multicolumn{3}{l}{
    $^a$\cite{Tillotson1962};
    $^b$\cite{Benz1999};
    $^c$\cite{Lundborg1967};
    $^d$\cite{Jutzi2008}.}
	\end{tabular}
	\label{table:model_parameters}
\end{table}
\FloatBarrier

The simulations were ran until $T$ = 500 seconds after the impact. To model these late times, after 15 seconds, we switch to a ``fast integration scheme'', as described in \cite{Jutzi2022, Raducan2022}. At this time, the initial shock and fragmentation phase are over, and the late-stage evolution is governed by low-velocity granular flow. This allows us to artificially change the material properties of the target to a low sound-speed medium, allowing for a larger timestep. In this calculation phase, we apply a simplified Tillotson equation of state (EoS) for all materials, in which all energy-related terms are set to zero. The remaining leading term of the EoS is governed by the bulk modulus, given by $P = A(\rho/\rho_0 - 1)$, which also determines the magnitude of the sound speed. We use $A\approx$ 1 MPa and also reduce the shear modulus proportionally.

\subsection{Largest remnant calculations}

In the collisional disruptions at the scales investigated here, the largest remnant is the accumulation of an array of gravitationally bound materials (i.e., intact monolithic fragments, re-accumulated dust, etc.). To quantify the mass of the largest remnant formed by the reaccumulation process of the smaller pieces, we use a ``fragment search'' iterative procedure introduced in \cite{Benz1999}.

This method identifies gravitationally bound aggregates by calculating the binding energy of all particles and fragments relative to either the largest intact fragment or, if too small, the particle closest to the potential minimum. Initially, this seed particle marks the nucleation point for the total bound mass. Unbound particles are removed, and the centre of mass position and velocity of the aggregate are computed. The process is repeated, recalculating the binding energy for the remaining fragments and particles relative to this new position and velocity, with unbound particles discarded at each iteration. Typically, convergence occurs within a few iterations, with few particles lost after the initial 2-3 steps. Finally, the method verifies that the members of this gravitationally bound aggregate are close spatially, using a friends-of-friends algorithm. Information such as mass, position, velocity, angular momentum, and moment of inertia is determined for this aggregate, consisting of smaller fragments and individual particles.

This method was shown to be accurate in determining the largest reaccumulated fragment ($M_{lr}$), as long as it has a significant size ($M_{lr}/M_{tot} >$ 10-20\%) \citep{Jutzi2010}, where $M_{tot}$ is the total initial mass of the asteroid \citep{Jutzi2010}.

\section{Results}

\subsection{Effects of the boulder packing}
\begin{figure}[ht]
\centering
\includegraphics[width=0.95\linewidth]{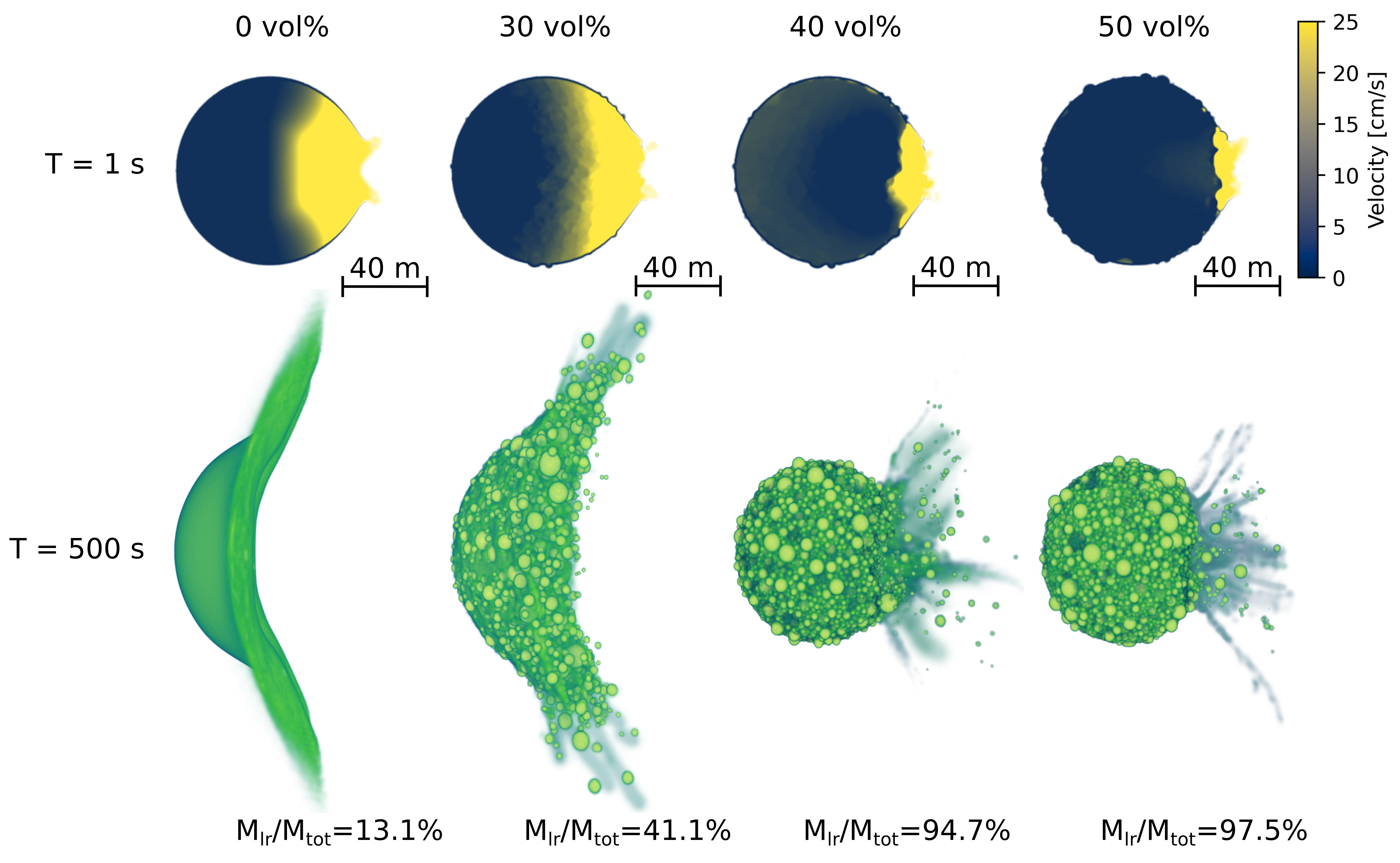}
\caption{Bern SPH simulations of collisions between a 87.90$\times$86.96$\times$57.16 m target and impactor of $R_p$ = 1.84 m ($M_p$ = 2.6$\times$10$^4$ kg), with a relative velocity of 6 km/s and a vertical impact angle. Targets with four different packing (0--50 vol\%) are investigated. (top row) Cross-section of SPH simulations showing the velocity field in the target at $T$ = 1 s. (bottom row) SPH simulations show the degree of disruption at $T$ = 500 s and the size of the largest remnant ($M_{lr}$). }
\label{fig:disrupt}
\end{figure}

We find that the interior structure of the rubble-pile asteroid plays a significant role in the outcome of the impact. Figure~\ref{fig:disrupt} shows the simulations outcome from a $M_p$ $\approx$ 2.6$\times$10$^4$ kg ($R_p$ = 1.84 m) impactor hitting four Dimorphos-like ellipsoidal targets (87.90$\times$86.96$\times$57.16 m), at 6 km/s. The boulder packing (i.e., volume occupied by large, $>$2.5 m, boulders) increases from left (homogeneous; 0 vol\%) to right (50 vol\%). For all target scenarios, the initial target mass is 4.15 $\times$ 10$^9$ $\pm$ 5\% kg. 

\FloatBarrier

The upper row in Figure~\ref{fig:disrupt} displays cross-sections of the four SPH simulations, showing the magnitude of the velocity field, at $T$ = 1 second after the impact. With increasing boulder packing, the pressure generated by shock wave propagation is attenuated by the presence of the boulders. In targets with higher boulder packing, a greater amount of energy is dissipated in the process of disrupting the boulders near the impact site, and the interlocking of boulders further hinders the shear motion of materials. Consequently, this leads to lower overall particle velocities within the target. This, in turn, results in less material being excavated by the impact and escaping the body's gravitational field. The largest remnant from impacting a $M_p$ = 2.6$\times$10$^4$ kg projectile on these rubble-pile targets with an initial mass, $M_{tot} \approx$ 4.15 $\times$ 10$^9$ $\pm$ 5\% kg, is $M_{lr}$ = 13.1\% of $M_{tot}$ for a homogeneous structure, $M_{lr}$ = 41.1\% for a target with 30 vol\% and $M_{lr}$ = 94.7\% and $M_{lr}$ = 97.5\% for 40 vol\% and 50 vol\%, respectively (Fig.~\ref{fig:disrupt}, bottom panels). This means that the same impact would excavate only a few percent of the target mass if the boulder packing is larger than about 40\%, but catastrophically disrupt the targets with lower boulder packing ($\lesssim$ 30 vol\%). 

\subsection{Catastrophic disruption threshold for vertical and oblique impacts}

\begin{figure}[ht]
\centering
\includegraphics[width=0.5\linewidth]{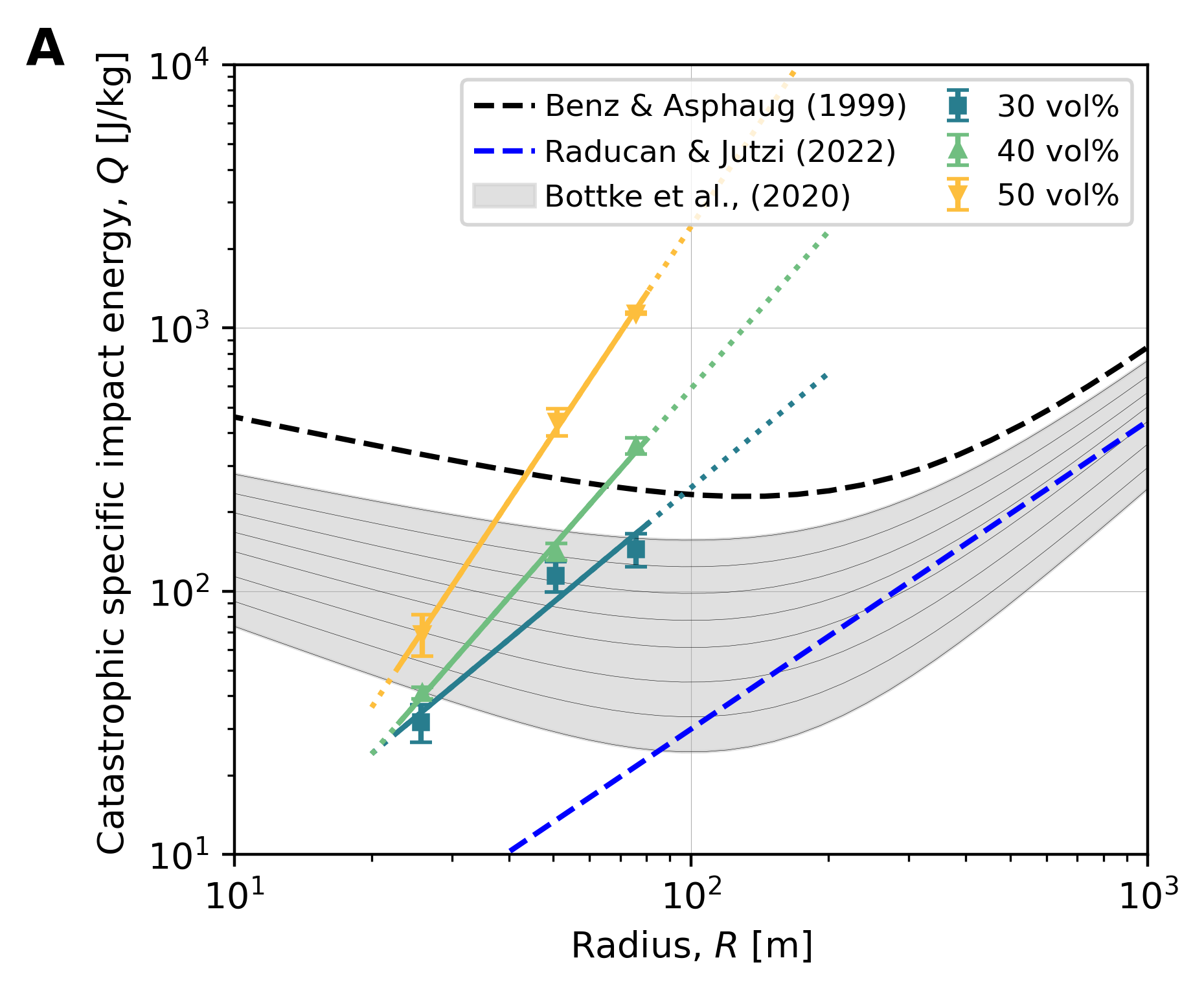}\includegraphics[width=0.5\linewidth]{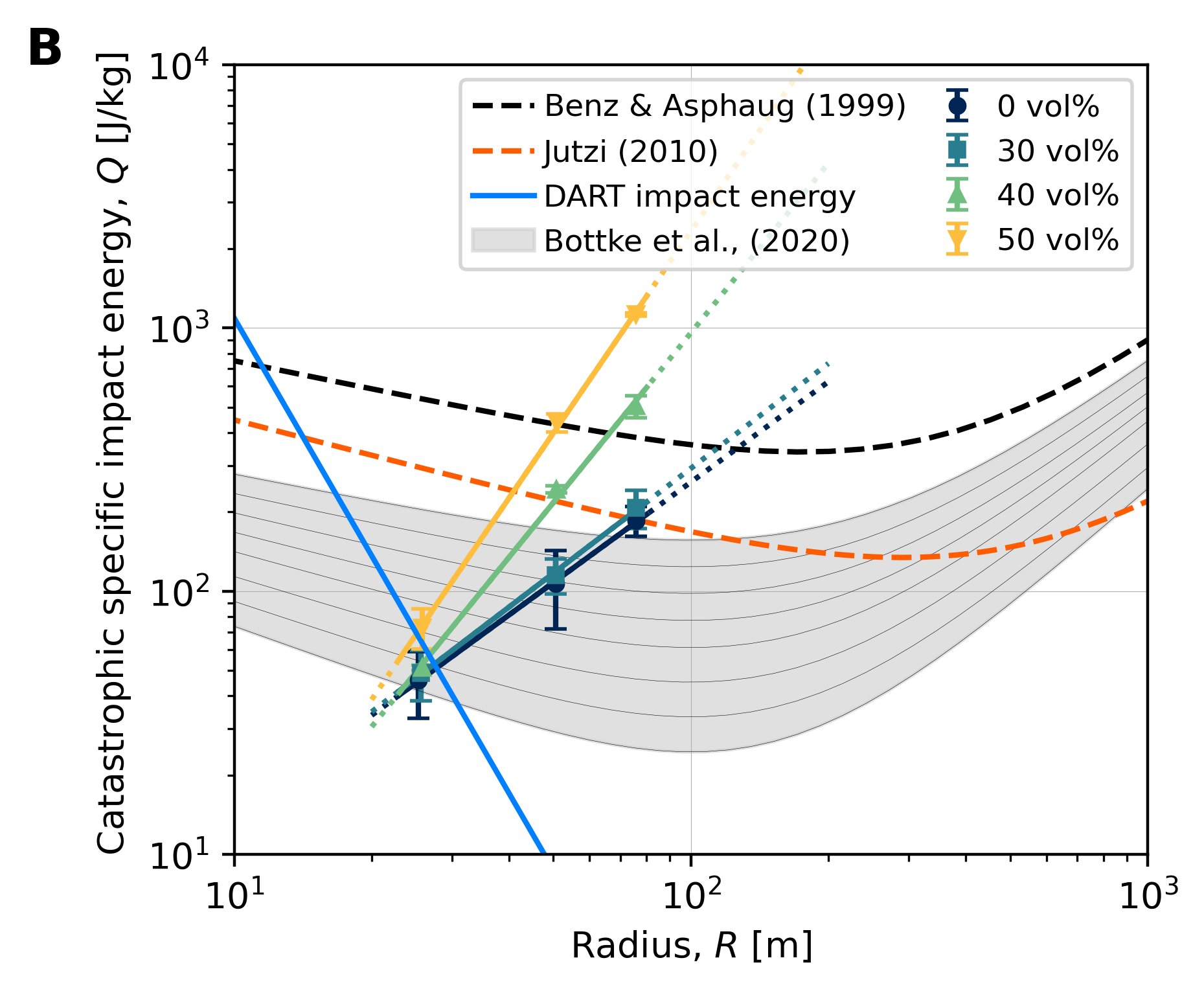}
\caption{Specific impact energy required for a catastrophic disruption per unit mass, $Q$, as a function of target radius, $R$, for targets with different boulder packing (0 vol\% to 50 vol\%). A) Vertical impact and B) Oblique, 45$^\circ$ impact. The impact location is the same for both impact angles (see the Numerical Methods section). The DART impact energy is plotted in both A and B. Results from \cite{Benz1999, Raducan2022, Jutzi2010} are plotted for comparison for the respective impact angle. $Q^*_D$ curves used by \cite{Bottke2020} are impact angle independent and are plotted in both A and B. }
\label{fig:qd_vert}
\end{figure}

From our simulations of vertical impacts into $\approx$ 150 m targets at 6 km/s, we find that with increasing boulder packing, an increasing amount of impact energy per unit mass is required to catastrophically disrupt the target (i.e., eject half of the initial mass). We find $Q^*_D$ = 80 $\pm$ 9 J/kg for a homogeneous target, $Q^*_D$ = 145 $\pm$ 21 J/kg for a 30 vol\% target, $Q^*_D$ = 357 $\pm$ 25 J/kg for a 40 vol\% target and $Q^*_D$ = 1136 $\pm$ 11 J/kg for a 50 vol\% target. Similar trends are obtained for the $\approx$ 100 m and $\approx$ 50 m ellipsoidal targets. Fig.~\ref{fig:qd_vert}a shows the
specific impact energy required for a catastrophic disruption, per unit mass, $Q$, as a function of target radius from our SPH simulations. 

In the gravity regime, when the gravitational force of the target dominates over the tensile strength of the body, $Q$ increases with target size, $R$, and impact velocity, $U$, as described by the scaling relationship \citep{Housen1990}, 
\begin{equation}\label{eq:qq}
    Q = a_g R^{3\mu_g}U^{2-3\mu_g},
\end{equation}
where $a_g$ is a constant, and $\mu_g$ is the coupling parameter to the target in the gravity regime. Using a best-fit algorithm, we find $a_g$ and $\mu_g$ (Table~\ref{table:constants}). The values for targets with a boulder packing lower than 30 vol\% have values comparable with what is generally assumed for typical porous materials ($\mu_g$ = 0.40–0.42, \citep{Jutzi2017}; $\mu_g$ = 0.33–0.36, \citep{Ballouz2020}). However, the derived $\mu_g$ for $>$30 vol\% exceeds these values, indicating other effects (boulder interlocking) are becoming more and more relevant with increasing boulder packing.

\begin{table}[ht]
\caption{Constants obtained from SPH simulations for catastrophic disruption threshold scaling relationships \citep{Housen1990}.}
\begin{tabular}{c@{\hskip 1.5cm}cc@{\hskip 1.5cm}cc}
 & \multicolumn{2}{c}{90$^\circ$ impact} & \multicolumn{2}{c}{{45$^\circ$ impact}} \\
Boulder packing & $a_g$ & $\mu_g$ & $a_g$ & $\mu_g$  \\
\toprule
0 vol\% & (2.15 $\pm$ 0.33) $\times$ 10$^{-4}$ & 0.34 $\pm$ 0.10 &   (1.29 $\pm$ 0.12) $\times$ 10$^{-3}$  & 0.42 $\pm$ 0.01 \\
30 vol\% & (2.55 $\pm$ 0.43) $\times$ 10$^{-3}$ & 0.48 $\pm$ 0.12 &  (1.83 $\times$ 0.18) $\times$ 10$^{-3}$   & 0.44 $\pm$ 0.02  \\
40 vol\% & (5.52 $\pm$ 0.38) $\times$ 10$^{-2}$ & 0.66 $\pm$ 0.04 &  (1.17 $\times$ 0.11) $\times$ 10$^{-1}$   & 0.71 $\pm$ 0.04 \\
50 vol\% & 2.96 $\pm$ 1.37 & 0.87 $\pm$ 0.03 &  2.17 $\pm$ 0.73  & 0.84 $\pm$ 0.02 \\              
\bottomrule

\end{tabular}
\label{table:constants}
\end{table}

Our results show that in the size range studied here (up to $R$ $\approx$ 100 m), small rubble-pile asteroids with a low boulder packing ($\lesssim$ 30 vol\%) are much easier to disrupt than monolithic bodies \citep{Benz1999,Jutzi2010}. $R$ $\approx$ 50 m with low boulder packing targets have a catastrophic impact energy ($Q$) about 4 times smaller than derived by \citep{Benz1999} for monolithic targets, while a $R$ $\approx$ 25 m has a $Q$ about 20 times smaller than a monolithic target \citep{Benz1999}. 

Asteroids with high packing of boulder also have smaller $Q$ at small sizes ($R$ $\lesssim$ 40 m), but are are much more resistant to impacts at $R$ $\gtrsim$ 80 m. To disrupt $R$ $\gtrsim$ 80 m targets, about 3.5 times more energy is required than to disrupt monolithic targets in the same size range \citep{Benz1999}. There are several contributing factors to these results. Firstly, the relatively high porosity (45\%) of the target attenuates the shock wave in comparison to a low-porosity monolithic target. At the same time, the interlocking of boulders inhibits shear motion, consequently diminishing the impact-induced velocity gradients within the target.

In contrast to the homogeneous and cohesionless targets studied in \cite{Raducan2022}, the rubble-pile targets studied here are harder to catastrophically disrupt. This can be attributed to the distinct material mechanical properties considered, particularly the initial porosity, grain density, and bulk modulus of the target material.

The same trends observed for vertical impacts are also observed for oblique, 45 $^\circ$, impacts. We find $Q^*_D$ = 185 $\pm$ 24 J/kg for a homogeneous target, $Q^*_D$ = 207 $\pm$ 34 J/kg for a 30 vol\% target, $Q^*_D$ = 504 $\pm$ 50 J/kg for a 40 vol\% target and $Q^*_D$ = 1125 $\pm$ 15 J/kg for a 50 vol\% target (Fig.~\ref{fig:disrupt}). The best-fit $a_g$ and $\mu_g$ constants for Eq.~\ref{eq:qq} are summarised in Table~\ref{table:constants}. 

Once more, in cases of high boulder packing, the slope of $Q$ (Eq.~\ref{eq:qq}) is notably steeper than what has been observed in monolithic and homogeneous targets \citep{Benz1999, Jutzi2010}. This indicates that also in the case of oblique impacts, small targets (with $R$ $\lesssim$ 40 m) are more prone to catastrophic disruption than previously anticipated, whereas larger targets ($R$ $\gtrsim$ 80 m) demonstrate greater resilience to impacts.

\section{Discussion}

\subsection{Deflection vs. disruption}

A catastrophic disruption event reduces the target to less than half of its original mass \citep[e.g.,][]{Holsapple2019} and leads to the creation of a cloud of potentially hazardous fragments from the ejecta. The trajectories of these fragments are highly sensitive to unpredictable variables, such as the asteroid's internal structure, introducing uncertainty in disruption outcomes. In contrast, deflection is a gradual process for adjusting the trajectory of an incoming NEO. Although subcatastrophic impacts such as DART produce significant ejecta \citep{Graykowski2023, Moreno2023, Jewitt2023, Roth-PSJ-2023}, the observed sizes of the ejected fragments are significantly smaller compared to disruption events mainly because of the sensitivities of telescopes observing such fragments. While these definitions distinguish between different outcomes, they also form a continuum within kinetic impactor technology and one spacecraft can achieve either outcome.

The deflection efficiency of a kinetic impactor depends on the target material properties and structure \citep{Raducan2019, Raducan2020, Raducan2022}, and it can be quantified in terms of a momentum enhancement factor, $\beta$. $\beta$ is defined by the momentum balance of the kinetic impact \citep{Cheng2023},
\begin{equation}
    M\Delta \boldsymbol{v} = m \boldsymbol{U} + m (\beta-1)(\boldsymbol{\hat{E}} \cdot \boldsymbol{U})\boldsymbol{\hat{E}}
\end{equation}
where $M$ is the mass of the asteroid, $\boldsymbol{U}$ is the impact velocity relative to the asteroid and $\boldsymbol{\hat{E}}$ is the net ejecta direction. The ejecta produced by the impact can enhance the momentum transfer efficiency. A value of $\beta \approx$ 1 would imply that the contribution of ejecta recoil to the momentum transfer was minimal. Conversely, a $\beta >$ 2 would indicate that the momentum from the ejecta exceeded that of the incident momentum from the kinetic impactor.
For DART, the observed period change \citep{Thomas2023} corresponds to a momentum enhancement factor, $\beta$, between 2.2 and 4.9, depending on the mass of Dimorphos \citep{Cheng2023}.

\begin{figure}[ht]
\centering
\includegraphics[width=0.5\linewidth]{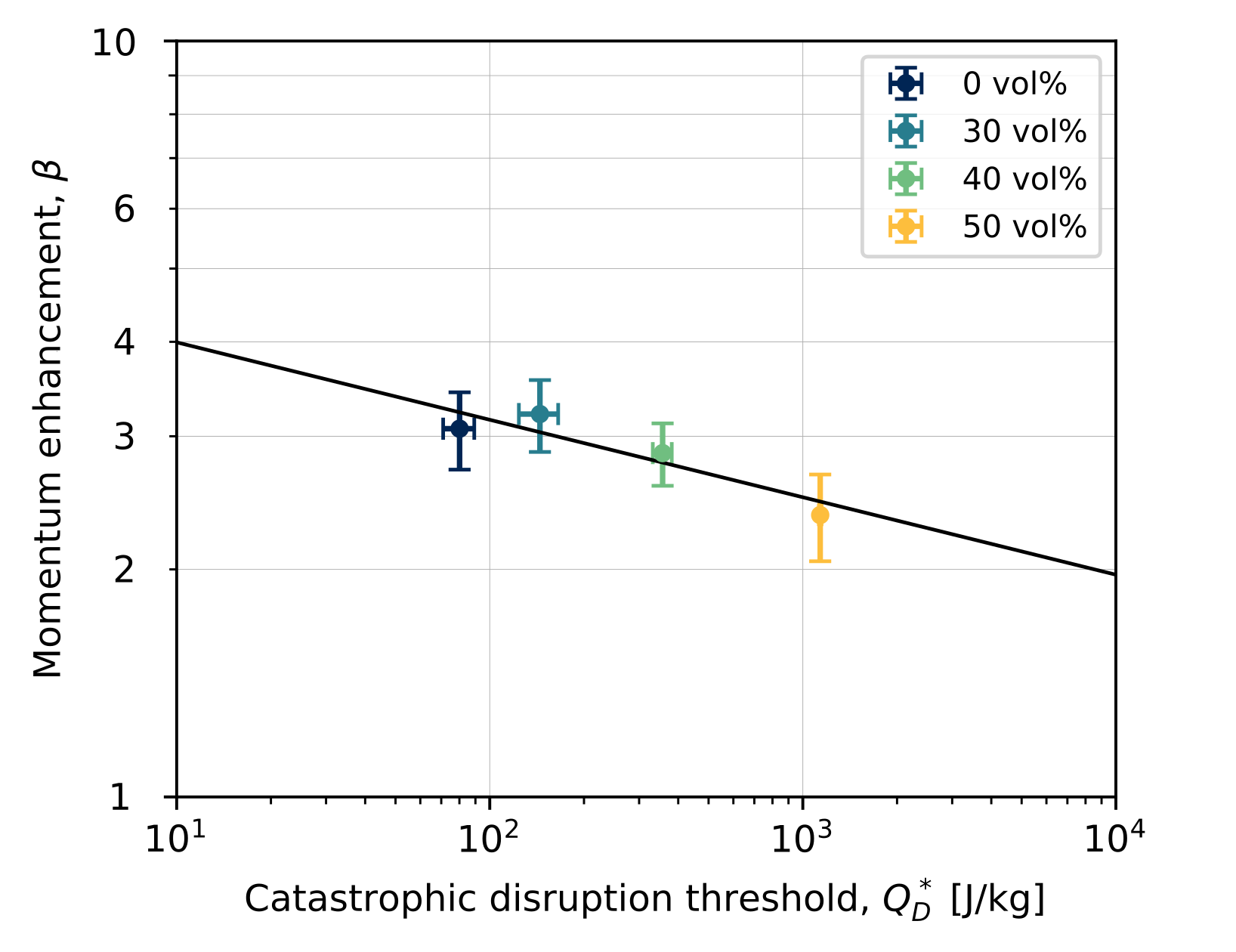}
\caption{Momentum enhancement, $\beta$ (from \cite{Raducan2023}) and the catastrophic disruption threshold, $Q^*_D$ (this study) for Dimorphos-sized ellipsoidal targets, with boulder packing between 0 vol\% and 50\%. $\beta$ is calculated from numerical simulations using the DART impact conditions (580 kg at 6 km/s) \citep{Raducan2023}. The observed change in the orbital period of Dimorphos, of $\approx$ 33 minutes \citep{Thomas2023}, corresponds to $\beta$ = 2.2 to 4.9, depending on the mass of the asteroid, which is still uncertain.}
\label{fig:beta}
\end{figure}

For an asteroid of a given size, the momentum enhancement ($\beta$) and the catastrophic disruption threshold ($Q^*_D$) are anti-correlated. Figure~\ref{fig:beta} shows $\beta$ values derived by \cite{Raducan2023}, as a function of $Q^*_D$ (from this study) for Dimorphos-sized ellipsoidal targets, with different boulder packing. The calculated $\beta$ values are for the DART impact conditions (580 kg at 6 km/s) and take into account the impact angle and impact location. Impacts on a target with a different curvature may result in a different $\beta$ value \citep{Hirabayashi2023}, however, the same trends are expected. We find that a target with no large boulders (0 vol\% packing) results in a large deflection (large $\beta$), however, it is also easier to catastrophically disrupt. On the other hand, for a target with a high boulder packing (50 vol\%), $\beta$ is reduced due to armouring and boulder interlocking \citep{Raducan2023}, and the asteroid is also significantly harder to catastrophically disrupt compared to a target with a low boulder packing. 

These findings imply that a greater momentum is necessary for an asteroid with a high $Q^*_D$ to achieve the desired deflection ($\Delta \boldsymbol{v}$). Therefore, deflecting an asteroid with a high boulder packing would require a larger and/or faster kinetic impactor. On the other hand,  in the case of an asteroid with a low $Q^*_D$, smaller impactors are preferable due to the heightened risk of disrupting the asteroid. Additionally, with a high $\beta$, less spacecraft momentum is needed to attain the desired $\Delta\boldsymbol{v}$. 
When it comes to smaller asteroids, accurately assessing their response to impacts becomes paramount. $Q^*_D$ becomes particularly critical, and its reliance on size underscores the need for precision in predictions. For such cases, the deflection approach must be tailored to the specific characteristics of the target asteroid. In the scenarios where the warning time is sufficiently long, a reconnaissance mission emerges as a prudent choice. This mission may incorporate a small impactor as part of its payload. This approach serves a dual purpose: not only does it offer an opportunity to assess the asteroid's impact response firsthand, but it also facilitates the estimation of crucial parameters like $\beta$ and $Q^*_D$. These assessments inform the deflection strategy. The deflection strategy for small asteroids is contingent on the anticipated response to impacts and can take several forms. Multiple small impacts may be employed, each contributing incrementally to the overall deflection. Alternatively, a single large high-energy impactor may be deployed, delivering a substantial momentum transfer. In more extreme cases, disruption may be the chosen method, potentially resulting in the fragmentation of the asteroid.


The highest risk of impacting Earth is posed by asteroids in the $\approx$ 20 to $\approx$ 90 meters diameter range. However, a mission aimed at testing the deflection strategies of near-Earth asteroids in this size range has yet to be undertaken. The impact of a kinetic impactor, similar to DART in both size and velocity at 6 km/s, may catastrophically disrupt an asteroid smaller than approximately 80 meters in diameter. As we continue to refine our deflection strategies, it becomes increasingly clear that any future missions must take the possibility of disruption into account, unless this outcome is expressly intended. This strategic foresight ensures that our efforts to defend against potential asteroid threats are both targeted and effective.

\subsection{Implications for the evolution of small asteroids}

Our investigation of the disruption threshold for various interior structures within rubble-pile asteroids shows that the catastrophic disruption threshold, $Q^*_D$, varies significantly contingent upon the boulder packing. In comparison with monolithic targets, $Q^*_D$ of rubble-pile targets is notably smaller for $R <$ 40 m and considerably larger for $R >$ 80 m. These findings not only bear consequences for planetary defence strategies, influencing how we may redirect or disrupt potential Earth-threatening asteroids but also affect our understanding of the age and structural evolution of rubble-pile asteroids.

Our findings indicate that small rubble-pile asteroids (with radii $R$ less than a few tens of meters) would withstand far fewer and less energetic collisions than their monolithic counterparts. Consequently, the collisional lifetime of rubble-pile asteroids in this size range is less than previously estimated by \cite{Bottke2020} (Fig.~\ref{fig:qd_vert}). This conclusion opens up the possibility of a broader spectrum in the composition of the small asteroid population, suggesting it may include both younger rubble-pile asteroids, whose lifespans are shorter due to their susceptibility to collisions, and older, more resilient monolithic asteroids. 

On the other hand, rubble-pile asteroids in the hundred of metre range (e.g., asteroids Dimorphos, Itokawa, Ryugu, Bennu) may be much older. We note that here we call ``age'' the period of time that has passed since the last disruption event (i.e., it corresponds to the `collisional lifetime'). The age of the surface (i.e., the time since the last resurfacing event) may be much younger, as shown in \cite{Raducan2022}. 

These new $Q^*_D$ trends found in this study could aid in interpreting the cratering record on these small asteroids and their relative ages, shedding new light on the dynamics and evolution of these bodies. This newfound insight calls for a re-evaluation of asteroid evolution models \citep[e.g.,][]{Bottke2020}, potentially adjusting the estimated lifetimes of small bodies in the asteroid belt and elsewhere in the Solar System.

\section{Conclusions}

We used the Bern SPH code to numerically model impact events into small, less than 150 m in diameter, rubble-pile asteroids, with varying boulder packing (between 0 and 50 vol\%). In our models, we used the insight gained from the DART impact on asteroid Dimorphos to calibrate the mechanical properties of the target material. Our findings highlight the important role of the interior structure in determining the impact's outcome and whether it leads to a subcatastrophic or catastrophic event. The same impact energy can catastrophically disrupt a target with a low boulder packing ($\lesssim$ 30 vol\%), while ejecting only a few percent of material from a target of the same mass, but high boulder packing ($\gtrsim$ 40 vol\%). 

We find that in the case of 50 m in diameter asteroids the catastrophic disruption threshold, $Q^*_D$ can be up to 20 times lower for rubble piles in comparison to monolithic targets. At the same time, targets larger than 160 m in diameter can require up to 3.5 times more energy to disrupt them, compared to monolithic targets. 

Our simulations indicate that a DART-sized spacecraft, at 6 km/s possesses the potential to cause catastrophic disruption in rubble-pile asteroids smaller than approximately 80 meters in diameter. Consequently, it is important for future deflection missions to carefully consider the potential for disruption, unless such an outcome is deliberately intended. Our result also implies that asteroids with diameters $\lesssim$ 50 m may be much younger than previously predicted while larger asteroids may be much older.

\section{Acknowledgements}
SDR and MJ acknowledge support from the Swiss National Science Foundation (project number 200021\_207359). This work was supported in part by the DART mission, NASA Contract \#80MSFC20D0004 to JHU/APL. PM acknowledges funding support from ESA and CNES.


\newpage
\bibliography{refs.bib} 
\bibliographystyle{aasjournal}

\newpage

\appendix

\section{Supplementary}

\begin{table}[ht]
\caption{Table of results from Bern SPH simulations of vertical impacts at 6 km/s into $\approx$ 150 m in diameter rubble pile targets.}
\begin{tabular}{cccccccccc}
 \multicolumn{3}{c}{Target} & & \multicolumn{4}{c}{Impactor}  & & Fraction \\
$R_t$$^*$ (m) & Packing & $M_t$ (kg) & & $R_p$ (m) & $M_p$ (kg) & $E_{kin}$/$E_{kin(DART)}$ & $Qr$ (J/kg) & & remaining (\%)\\
\hline
\multirow{6}{*}{75.82} & \multirow{6}{*}{0 vol\%} & \multirow{6}{*}{3.25$\times$10$^9$} && 0.62 & 9.76$\times$ 10$^2$ & 1.59 & 5.41 & & 96.15\% \\

                       &  &  & & 0.78 & 1.94$\times$10$^3$ & 3.16 & 10.74 & & 92.18\% \\
                       &  &  & & 1.06 & 4.88$\times$10$^3$ & 7.94 & 27.03 & & 82.81\% \\
                       &  &  & & 1.24 & 7.81$\times$10$^3$ & 12.71 & 43.26 & & 71.46\% \\
                       &  &  & & 1.42 & 1.17$\times$10$^4$ & 19.08 & 64.80 & & 61.89\% \\
                       &  &  & & 1.84 & 2.51$\times$10$^4$ & 40.92 & 139.01 & & 13.13\% \\
\hline 

\multirow{6}{*}{75.83} & \multirow{6}{*}{30 vol\%} & \multirow{6}{*}{3.91$\times$10$^9$} && 0.62 & 9.76$\times$ 10$^2$& 1.59 & 4.49 & & 97.68\% \\

                       &  &  & & 0.78 & 1.94$\times$10$^3$ & 3.16 & 8.93 & & 95.19\% \\
                       &  &  & & 1.06 & 4.88$\times$10$^3$ & 7.94 & 22.47 & & 85.94\% \\
                       &  &  & & 1.24 & 7.81$\times$10$^3$ & 12.71 & 35.95 & & 74.83\%\\
                       &  &  & & 1.42 & 1.17$\times$10$^4$ & 19.08 & 53.86 & & 66.05\% \\
                       &  &  & & 1.84 & 2.51$\times$10$^4$ & 40.92 & 115.55 & & 41.09\% \\
\hline 

\multirow{9}{*}{75.91} & \multirow{9}{*}{40 vol\%} & \multirow{9}{*}{4.25$\times$10$^9$} && 0.62 & 9.76$\times$ 10$^2$ & 1.59 & 4.13 & & 99.55\% \\

                       &  &  & & 0.78 & 1.94$\times$10$^3$ & 3.16 & 8.22 & & 99.31\% \\
                       &  &  & & 1.06 & 4.88$\times$10$^3$ & 7.94 & 20.67 & & 98.85\% \\
                       &  &  & & 1.24 & 7.81$\times$10$^3$ & 12.71 & 33.08 & & 98.34\% \\
                       &  &  & & 1.42 & 1.17$\times$10$^4$ & 19.08 & 49.55 & & 96.38\% \\
                       &  &  & & 1.84 & 2.51$\times$10$^4$ & 40.92 & 106.30 & & 94.75\% \\
                       &  &  & & 2.44 & 5.88$\times$10$^4$ & 95.67 & 249.03 & & 79.49\% \\
                       &  &  & & 2.84 & 9.27$\times$10$^4$ & 150.86 & 392.59 & & 33.33\% \\
                       &  &  & & 3.12 & 1.23$\times$10$^5$ & 200.03 & 520.91 & & 1.53\% \\
\hline 

\multirow{9}{*}{75.92} & \multirow{9}{*}{50 vol\%} & \multirow{9}{*}{4.38$\times$10$^9$} && 0.62 & 9.76$\times$ 10$^2$ & 1.59 & 4.01 & & 99.93\% \\

                       &  &  & & 0.78 & 1.94$\times$10$^3$ & 3.16 & 7.97 & & 99.84\% \\
                       &  &  & & 1.06 & 4.88$\times$10$^3$ & 7.94 & 20.05 & & 99.48\% \\
                       &  &  & & 1.24 & 7.81$\times$10$^3$ & 12.71 & 32.10 & & 99.32\% \\
                       &  &  & & 1.42 & 1.17$\times$10$^4$ & 19.08 & 48.08 & & 98.55\% \\
                       &  &  & & 1.84 & 2.51$\times$10$^4$ & 40.92 & 103.15 & & 97.46\% \\
                       &  &  & & 2.84 & 9.27$\times$10$^4$ & 150.86 & 380.94 & & 94.51\% \\
                       &  &  & & 3.62 & 1.92$\times$10$^5$ & 315.42 & 788.97 & & 69.68\%\\
                       &  &  & & 4.42 & 3.50$\times$10$^4$ & 568.70 & 1487.43 & & 26.05\% \\
\bottomrule

\end{tabular}
$^*$Volume-equivalent target radius.
\end{table}

\begin{figure*}[ht]
\centering
\includegraphics[width=0.9\linewidth]{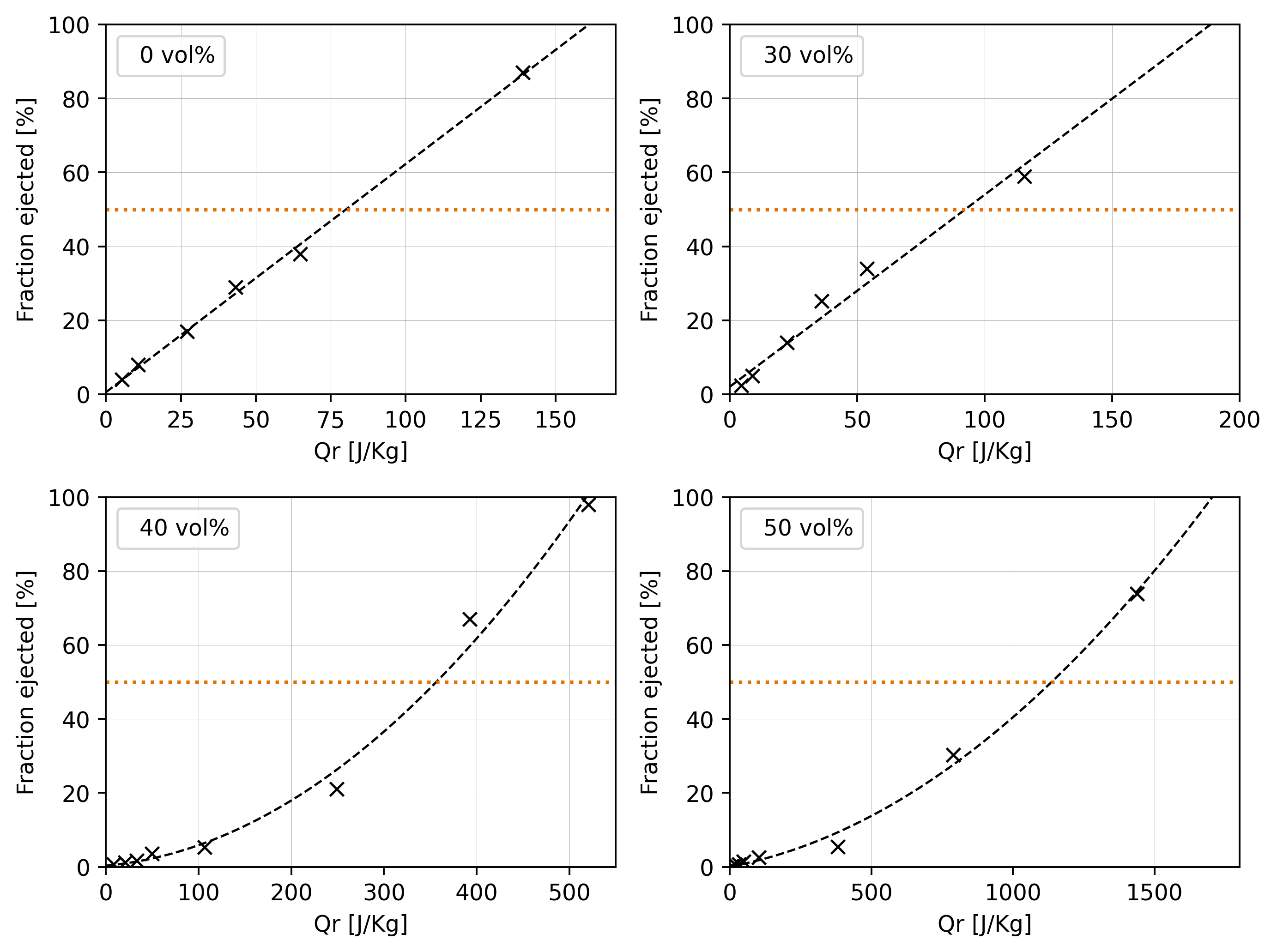}
\caption{Fraction of ejected material as a function of specific impact energy ($Q_r$), for 6 km/s impacts into $\approx$150 m rubble pile asteroids with varying boulder packing (0 to 50 \%). The intersection between the trend through the simulation data (denoted by the dashed lines) and the 50\% fraction ejected (denoted by the dotted lines) denotes the $Q^*_D$ for the specific target. }
\label{fig:qr_vertical}
\end{figure*}

\clearpage

\begin{table}[ht]
\caption{Table of results from Bern SPH simulations of oblique, 45 degrees impacts at 6 km/s into $\approx$ 150 m in diameter rubble pile targets.}
\begin{tabular}{cccccccccc}
 \multicolumn{3}{c}{Target} & & \multicolumn{4}{c}{Impactor}  & & Fraction \\
$R_t$$^*$ (m) & $Packing$ & $M_t$ (kg) & & $R_p$ (m) & $M_p$ (kg) & $E_{kin}$/$E_{kin(DART)}$ & $Qr$ (J/kg) &  & remaining (\%)\\
\hline
\multirow{3}{*}{75.82} & \multirow{3}{*}{0 vol\%} & \multirow{3}{*}{3.25$\times$10$^9$} && 1.42 & $1.17\times$10$^4$ & 19.3 & 64.80 & & 76.19\% \\

                       &  &  & & 1.84 & $2.51\times$10$^4$ & 41.4 & 139.01 & & 55.29\% \\
                       &  &  & & 2.44 & $5.88\times$10$^4$ & 96.5 & 325.65 & & 20.12\% \\
\hline 

\multirow{3}{*}{75.83} & \multirow{3}{*}{30 vol\%} & \multirow{3}{*}{3.91$\times$10$^9$} && 1.42 & $1.17\times$10$^4$& 19.3 & 53.86 & & 83.64\% \\

                       &  &  & & 1.84 & $2.51\times$10$^4$ & 41.4 & 115.55 & & 64.02\% \\
                       &  &  & & 2.44 & $5.88\times$10$^4$ & 96.5 & 270.68 & & 38.57\% \\
\hline 

\multirow{3}{*}{75.91} & \multirow{3}{*}{40 vol\%} & \multirow{3}{*}{4.25$\times$10$^9$} && 2.44 & $5.88\times$10$^4$ & 96.4 & 249.03 & & 88.40\% \\

                       &  &  & & 2.84 & $9.27\times$10$^4$ & 152.6 & 392.59 & & 63.33\% \\
                       &  &  & & 3.12 & $1.23\times$10$^4$ & 202.3 & 520.91 & & 41.26\% \\
\hline 

\multirow{3}{*}{75.92} & \multirow{3}{*}{50 vol\%} & \multirow{3}{*}{4.38$\times$10$^9$} && 2.84 & $9.27\times$10$^4$ & 152.6 & 380.94 & & 78.40\% \\   
                       &  &  & & 4.42 & $3.50\times$10$^4$ & 575.1 & 1436.20 & & 40.01\% \\
                       &  &  & & 4.92 & $4.82\times$10$^4$ & 793.2 & 1980.39 & & 11.87\% \\
\bottomrule

\end{tabular}

$^*$Volume-equivalent target radius.
\end{table}

\begin{figure}[ht]
\centering
\includegraphics[width=0.8\linewidth]{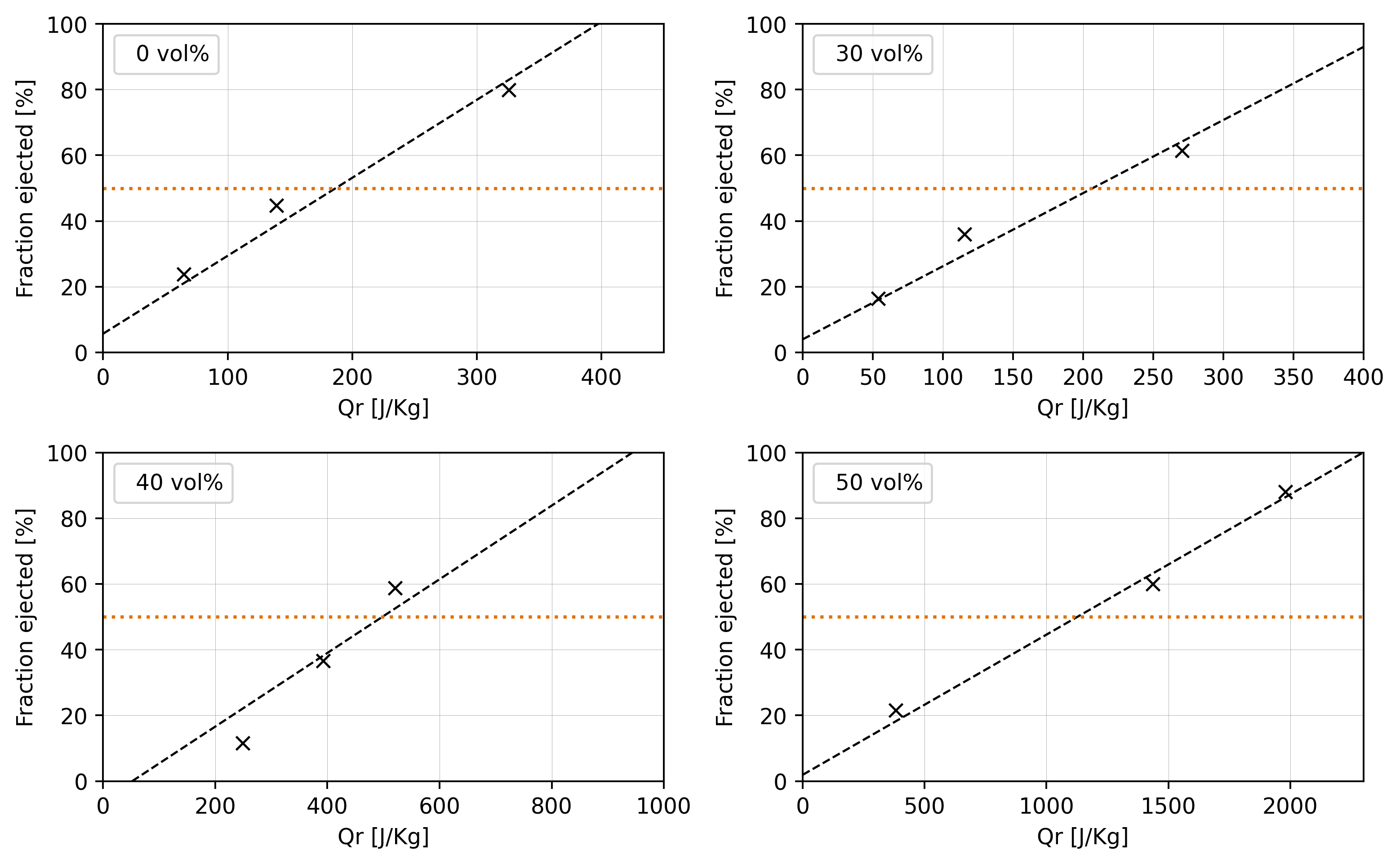}
\caption{Same as in Fig.~\ref{fig:qr_vertical} but for oblique (45 degrees), 6 km/s impacts into $\approx$ 150 m rubble pile asteroids with varying boulder packing (0 to 50 \%).}
\label{fig:qr_150_45}
\end{figure}

\FloatBarrier

\clearpage

\begin{table}[ht]
\caption{Table of results from Bern SPH simulations of vertical impacts at 6 km/s into $\approx$ 100 m in diameter rubble pile targets.}
\begin{tabular}{cccccccccc}
 \multicolumn{3}{c}{Target} & & \multicolumn{4}{c}{Impactor}  & & Fraction \\
$R_t$$^*$ (m) & $Packing$ & $M_t$ (kg) & & $R_p$ (m) & $M_p$ (kg) & $E_{kin}$/$E_{kin(DART)}$ & $Qr$ (J/kg) &  & remaining (\%)\\
\hline
\multirow{2}{*}{50.57} & \multirow{2}{*}{0 vol\%} & \multirow{2}{*}{9.64$\times$10$^8$} && 0.78 & $1.94\times$10$^3$ & 3.20 & 36.22 & & 79.94\% \\

                       &  &  & & 1.06 & $4.88\times$10$^3$ & 8.03 & 91.42 & & 35.75\% \\
\hline 

\multirow{2}{*}{50.58} & \multirow{2}{*}{30 vol\%} & \multirow{2}{*}{1.16$\times$10$^9$} && 1.06 & $4.88\times$10$^3$& 8.03 & 75.72 & & 69.87\% \\

                       &  &  & & 1.24 & $7.81\times$10$^3$ & 12.85 & 121.19 & & 45.54\% \\
\hline 

\multirow{2}{*}{50.75} & \multirow{2}{*}{40 vol\%} & \multirow{2}{*}{1.27$\times$10$^9$} && 1.06 & $4.88\times$10$^3$ & 8.03 & 69.16 & & 87.97\% \\

                       &  &  & & 1.42 & $1.17\times$10$^4$ & 19.25 & 165.82 & & 36.27\% \\
\hline 

\multirow{2}{*}{50.90} & \multirow{2}{*}{50 vol\%} & \multirow{2}{*}{1.32$\times$10$^9$} && 1.84 & $2.51\times$10$^4$ & 41.38 & 342.26 & & 53.88\% \\   
                       &  &  & & 2.44 & $5.88\times$10$^4$ & 96.75 & 801.75 & & 14.87\% \\
\bottomrule

\end{tabular}
$^*$Volume-equivalent target radius.
\end{table}

\begin{figure}[ht]
\centering
\includegraphics[width=0.9\linewidth]{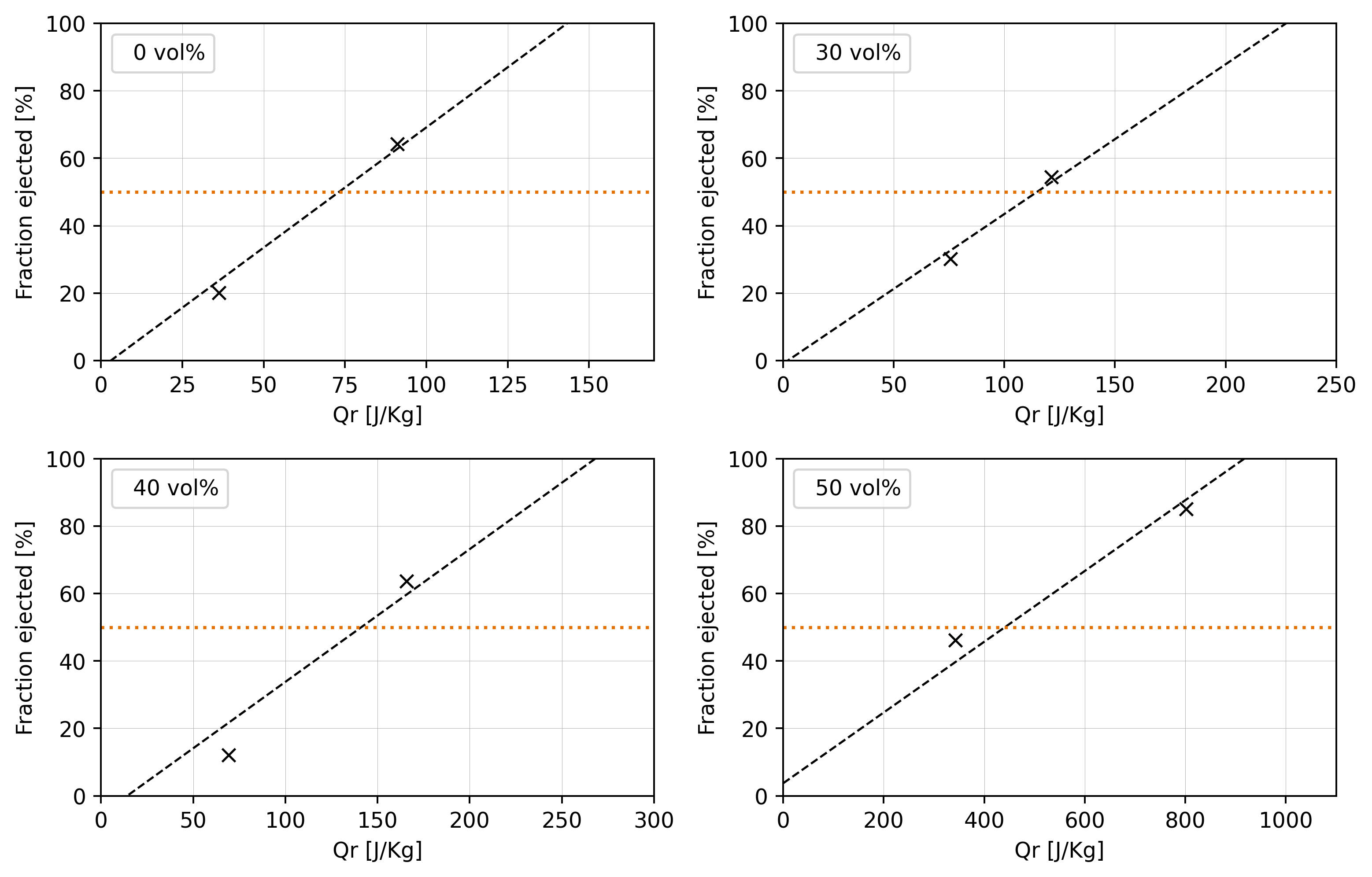}
\caption{Same as in Fig.~\ref{fig:qr_vertical} but for vertical, 6 km/s impacts into $\approx$110 m rubble pile asteroids with varying boulder packing (0 to 50 \%).}
\label{fig:qr_110}
\end{figure}

\FloatBarrier

\clearpage

\begin{table}[ht]
\caption{Table of results from Bern SPH simulations of oblique, 45 degrees impacts at 6 km/s into $\approx$ 100 m in diameter rubble pile targets.}
\begin{tabular}{cccccccccc}
 \multicolumn{3}{c}{Target} & & \multicolumn{4}{c}{Impactor}  & & Fraction \\
$R_t$$^*$ (m) & $Packing$ & $M_t$ (kg) & & $R_p$ (m) & $M_p$ (kg) & $E_{kin}$/$E_{kin(DART)}$ & $Qr$ (J/kg) &  & remaining (\%)\\
\hline
\multirow{3}{*}{55.57} & \multirow{3}{*}{0 vol\%} & \multirow{3}{*}{9.64$\times$10$^8$} && 0.78 & $1.94\times$10$^3$ & 3.20 & 36.22 & & 72.62\% \\

                       &  &  & & 1.06 & $4.88\times$10$^3$ & 8.03 & 91.42 & & 47.60\% \\
                       &  &  & & 1.42 & $1.17\times$10$^4$ & 19.30 & 218.46 & & 11.56\% \\
\hline 

\multirow{3}{*}{55.58} & \multirow{3}{*}{30 vol\%} & \multirow{3}{*}{1.16$\times$10$^9$} && 0.78 & $1.94\times$10$^3$& 3.20 & 30.13 & & 83.01\% \\

                    &  &  & & 1.06 & $4.88\times$10$^3$ & 8.03 & 75.72 & & 61.73\% \\
                    &  &  & & 1.42 & $1.17\times$10$^4$ & 19.30 & 181.55 & & 17.57\% \\
\hline 

\multirow{3}{*}{55.74} & \multirow{3}{*}{40 vol\%} & \multirow{3}{*}{1.27$\times$10$^9$} && 1.06 & $4.88\times$10$^3$ & 8.03 & 69.16 & & 98.05\% \\

                       &  &  & & 1.42 & $1.17\times$10$^4$ & 19.25 & 165.82 & & 77.58\% \\
                       &  &  & & 1.84 & $2.51\times$10$^4$ & 41.38 & 355.73 & & 17.57\% \\
\hline 

\multirow{3}{*}{55.90} & \multirow{3}{*}{50 vol\%} & \multirow{3}{*}{1.32$\times$10$^9$} && 1.06 & $4.88\times$10$^3$ & 8.03 & 342.26 & & 98.64\% \\   
                       &  &  & & 1.42 & $1.17\times$10$^4$ & 19.30 & 159.54 & & 88.05\% \\
                       &  &  & & 1.84 & $2.51\times$10$^4$ & 41.38 & 342.26 & & 59.42\% \\
\bottomrule

\end{tabular}
$^*$Volume-equivalent target radius.
\end{table}

\begin{figure}[ht]
\centering
\includegraphics[width=0.8\linewidth]{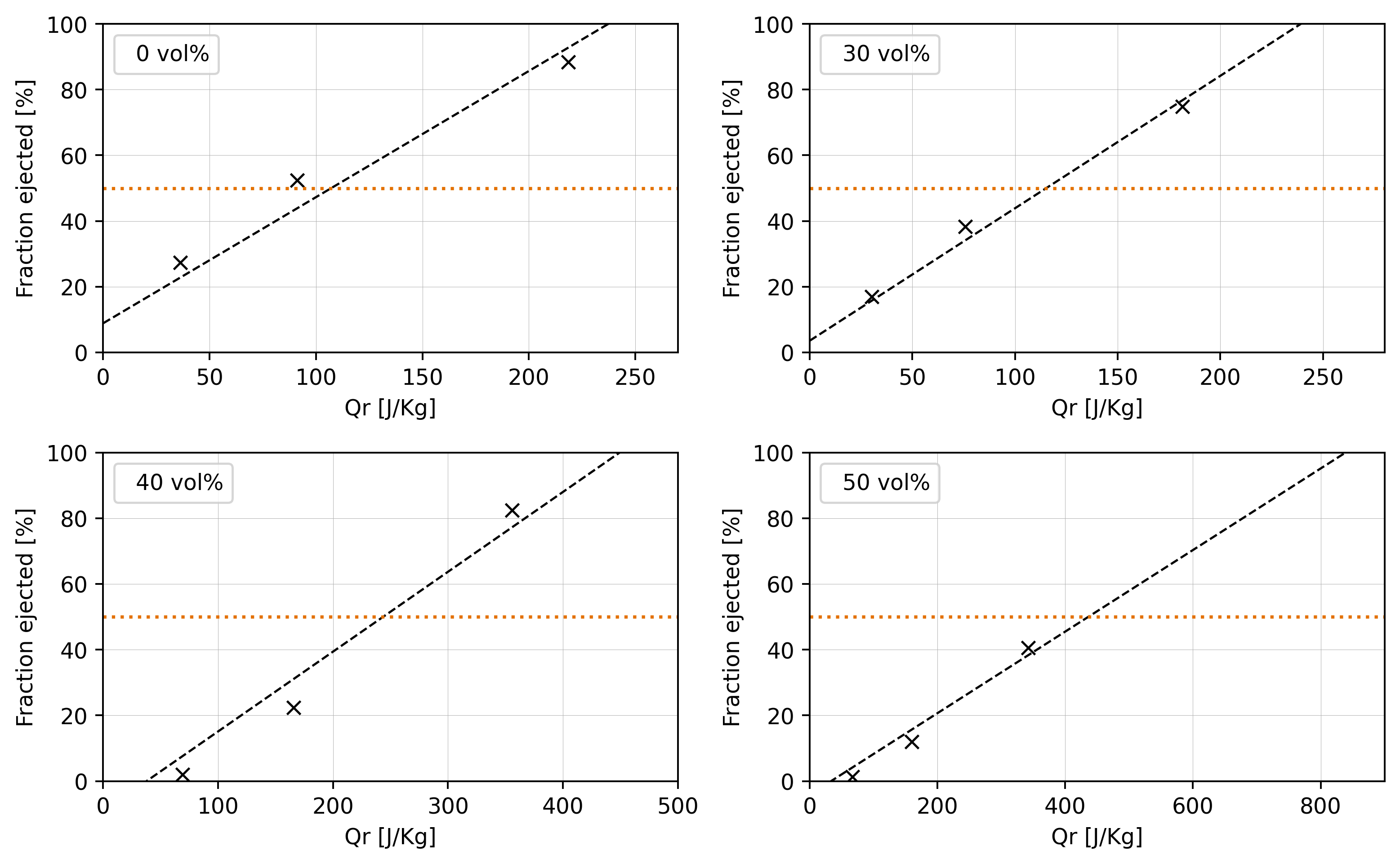}
\caption{Same as in Fig.~\ref{fig:qr_vertical} but for oblique (45 degrees), 6 km/s impacts into $\approx$ 100 m rubble pile asteroids with varying boulder packing (0 to 50 \%).}
\label{fig:qr_110_45}
\end{figure}

\clearpage

\begin{table}[ht]
\caption{Table of results from Bern SPH simulations of vertical impacts at 6 km/s into $\approx$ 50 m in diameter rubble pile targets.}
\begin{tabular}{cccccccccc}
 \multicolumn{3}{c}{Target} & & \multicolumn{4}{c}{Impactor}  & & Fraction \\
$R_t$$^*$ (m) & $Packing$ & $M_t$ (kg) & & $R_p$ (m) & $M_p$ (kg) & $E_{kin}$/$E_{kin(DART)}$ & $Qr$ (J/kg) &  & remaining (\%)\\
\hline
\multirow{3}{*}{25.32} & \multirow{3}{*}{0 vol\%} & \multirow{3}{*}{1.21$\times$10$^8$} && 0.36 & $1.79\times$10$^2$ & 0.29 & 26.67 & & 44.12\% \\

                       &  &  & & 0.42 & $2.84\times$10$^2$ & 0.46 & 42.35 & & 21.39\% \\
                       &  &  & & 0.48 & $4.25\times$10$^2$  & 0.69 & 63.22 & & 0.30\% \\
\hline 

\multirow{3}{*}{25.63} & \multirow{3}{*}{30 vol\%} & \multirow{3}{*}{1.51$\times$10$^8$} && 0.36 & $1.79\times$10$^2$  & 0.29 & 21.37 & & 73.92\% \\
                       &  &  & & 0.42 & $2.84\times$10$^2$ & 0.46 & 33.93 & & 42.24\% \\
                       &  &  & & 0.48 & $4.25\times$10$^2$  & 0.69 & 50.66 & & 18.93\% \\
\hline 

\multirow{2}{*}{25.80} & \multirow{2}{*}{40 vol\%} & \multirow{2}{*}{1.67$\times$10$^8$} && 0.42 & $2.84\times$10$^2$ & 0.46 & 30.68 & & 64.95\% \\

                       &  &  & & 0.54 & $6.05\times$10$^2$ & 0.98 & 65.21 & & 19.24\% \\
\hline 

\multirow{2}{*}{25.81} & \multirow{2}{*}{50 vol\%} & \multirow{2}{*}{1.72$\times$10$^8$} && 0.42 & $2.84\times$10$^2$ & 0.46 & 29.79 & & 92.86\% \\   
                       &  &  & & 0.54 & $6.05\times$10$^2$ & 0.98 & 63.31 & & 49.17\% \\
\bottomrule

\end{tabular}
$^*$Volume-equivalent target radius.
\end{table}

\begin{figure}[ht]
\centering
\includegraphics[width=0.8\linewidth]{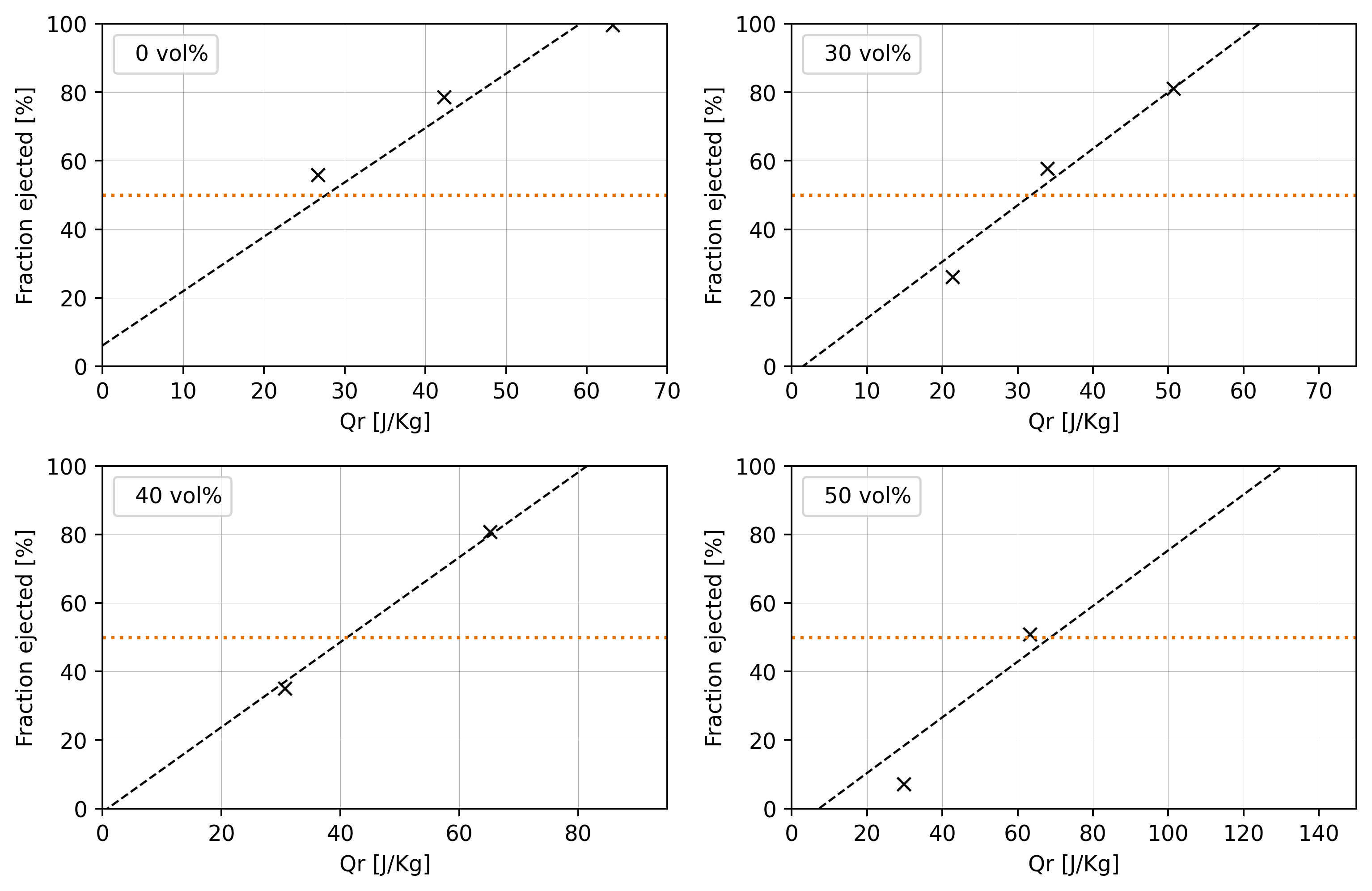}
\caption{Same as in Fig.~\ref{fig:qr_vertical} but for vertical, 6 km/s impacts into $\approx$ 50 m rubble pile asteroids with varying boulder packing (0 to 50 \%).}
\label{fig:qr_60}
\end{figure}

\clearpage

\begin{table}[ht]
\caption{Table of results from Bern SPH simulations of oblique, 45 degrees impacts at 6 km/s into $\approx$ 50 m in diameter rubble pile targets.}
\begin{tabular}{cccccccccc}
 \multicolumn{3}{c}{Target} & & \multicolumn{4}{c}{Impactor}  & & Fraction (\%) \\
$R_t$$^*$ (m) & $Packing$ & $M_t$ (kg) & & $R_p$ (m) & $M_p$ (kg) & $E_{kin}$/$E_{kin(DART)}$ & $Qr$ (J/kg) &  & remaining (\%)\\
\hline
\multirow{2}{*}{25.32} & \multirow{2}{*}{0 vol\%} & \multirow{2}{*}{1.21$\times$10$^8$} && 0.42 & $2.84\times$10$^2$ & 0.46 & 42.35 & & 41.52\% \\

                       &  &  & & 0.54 & $4.25\times$10$^2$  & 0.69 & 89.99 & & 13.47\% \\
\hline 

\multirow{2}{*}{25.63} & \multirow{2}{*}{30 vol\%} & \multirow{2}{*}{1.51$\times$10$^8$} && 0.42 & $2.84\times$10$^2$ & 0.46 & 33.93 & & 54.06\% \\

                       &  &  & & 0.54 & $4.25\times$10$^2$ & 0.69 & 72.12 & & 33.13\% \\
\hline 

\multirow{2}{*}{25.80} & \multirow{2}{*}{40 vol\%} & \multirow{2}{*}{1.67$\times$10$^8$} && 0.42 & $2.84\times$10$^2$ & 0.46 & 30.68 & & 68.73\% \\

                       &  &  & & 0.54 & $6.05\times$10$^2$ & 0.98 & 65.21 & & 36.92\% \\
\hline 

\multirow{2}{*}{25.81} & \multirow{2}{*}{50 vol\%} & \multirow{2}{*}{1.72$\times$10$^8$} && 0.42 & $2.84\times$10$^2$ & 0.46 & 29.79 & & 94.68\% \\   
                       &  &  & & 0.54 & $6.05\times$10$^2$ & 0.98 & 63.31 & & 52.02\% \\
\bottomrule

\end{tabular}
$^*$Volume-equivalent target radius.
\end{table}

\begin{figure}[ht]
\centering
\includegraphics[width=0.85\linewidth]{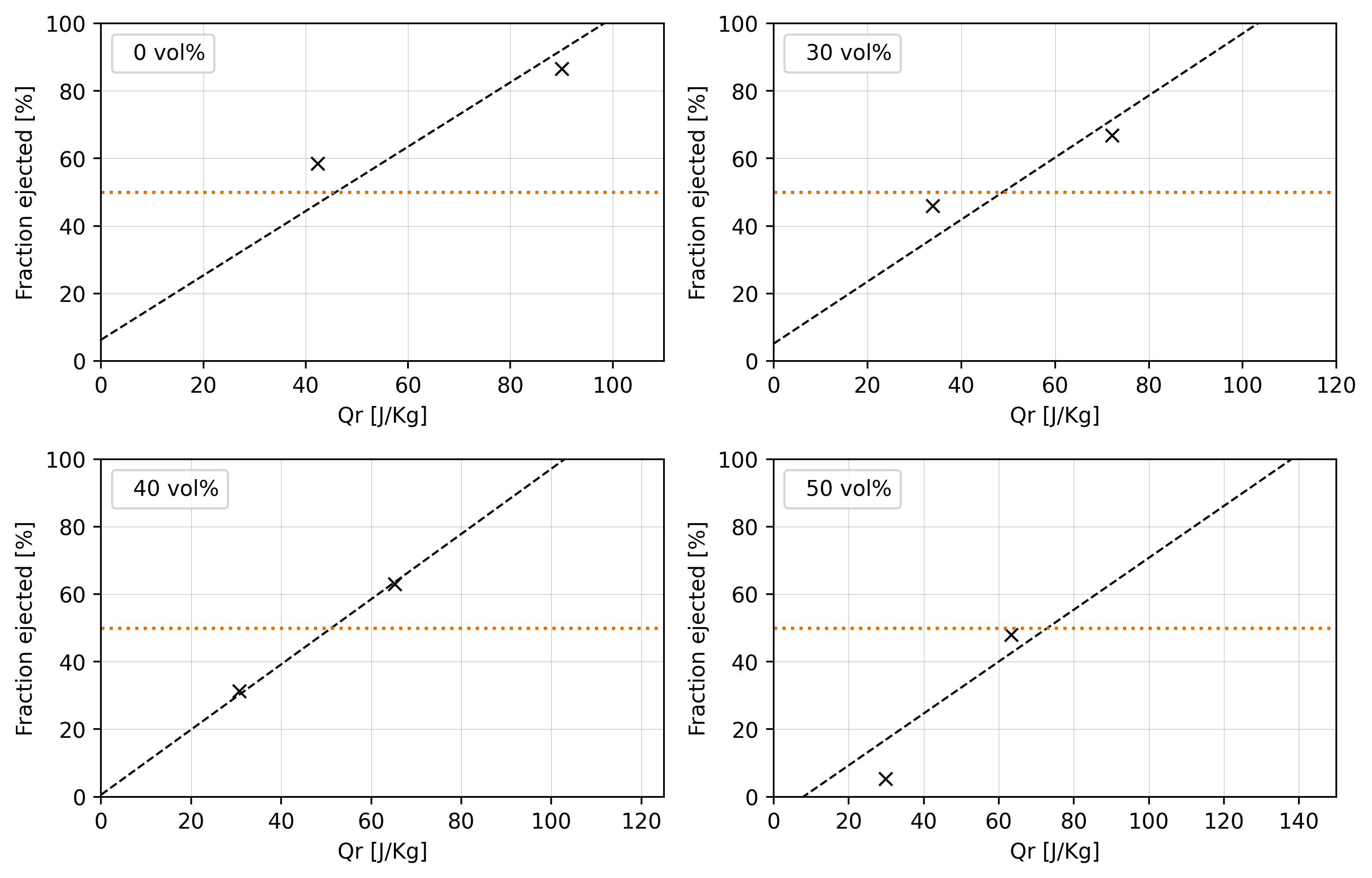}
\caption{Same as in Fig.~\ref{fig:qr_vertical} but for 45$^\circ$, 6 km/s impacts into $\approx$50 m rubble pile asteroids with varying boulder packing (0 to 50 \%).}
\label{fig:qr_60_45}
\end{figure}

\end{document}